\documentclass[runningheads,a4paper]{llncs}

\usepackage{amssymb}
\usepackage{amsfonts}
\usepackage{amsmath}
\usepackage{latexsym,epsfig}
\usepackage{graphicx} 
\usepackage[usenames,dvipsnames]{color} 
\usepackage{url} 
\usepackage{soul} 
\usepackage{array} 
\usepackage{verbatim}
\usepackage{hyperref}
\usepackage[ruled]{algorithm2e}
\usepackage{url}

\newcommand{\R}{\mathbb{R}}
\newcommand{\bigoh}{\mathcal{O}}

\newcommand{\keywords}[1]{\par\addvspace\baselineskip
\noindent\keywordname\enspace\ignorespaces#1}

\begin{document}

\mainmatter  
\title{Fitting Voronoi Diagrams to Planar Tesselations}

\titlerunning{Fitting Voronoi Diagrams}

\author{Greg Aloupis\inst{1} \and
Hebert P\'erez-Ros\'es\inst{2} \and 
Guillermo Pineda-Villavicencio\inst{3} \and\\
Perouz Taslakian\inst{4} \and 
Dannier Trinchet-Almaguer\inst{5}
}  
\authorrunning{Aloupis et al.}

\institute{Charg\'{e} de Recherches FNRS, Universit\'{e} Libre de Bruxelles, Belgium\\
\email{aloupis.greg@gmail.com}
\and
Department of Mathematics, University of Lleida, Spain,\\
and Conjoint Fellow, University of Newcastle, Australia\\
\email{hebert.perez@matematica.udl.cat}
\and
Center for Informatics and Applied Optimization, University of Ballarat, Australia,\\
and Department of Mathematics, Ben-Gurion University of the Negev, Israel\\
\email{work@guillermo.com.au}
\and
School of Science and Engineering, American University of Armenia\\
\email{ptaslakian@aua.am}
\and
AlessTidyCraft Software Solutions, Havana, Cuba\\
\email{trinchet@gmail.com}
}

\toctitle{Lecture Notes in Computer Science}
\tocauthor{Authors' Instructions}
\maketitle

\begin{abstract}
Given a tesselation of the plane, defined by a planar straight-line graph $G$, we want to find a minimal set $S$ of points in the plane, such that the Voronoi diagram associated with $S$ \lq fits\rq \ $G$. This is the Generalized Inverse Voronoi Problem (GIVP), defined in \cite{Trin07} and rediscovered recently in \cite{Baner12}. Here we give an algorithm that solves this problem with a number of points that is linear in the size of $G$, assuming that the smallest angle in $G$ is constant.  
\end{abstract}

\section{Introduction}
\label{intro}

Any planar straight-line graph (PSLG) subdivides the plane into cells, some of which may be unbounded. The Voronoi diagram (also commonly referred to as \emph{Dirichlet tesselation}, or \emph{Thiessen polygon})  of a set $S$  of $n$ points is a PSLG with $n$ cells, where each cell belongs to one point from $S$ and consists of all points in the plane that are closer to that point than to any other in $S$.

Let $G$ be a given PSLG, whose cells can be considered bounded and convex for all practical purposes. Indeed, if some cell is not convex, it can always be partitioned into convex subcells, thus yielding a finer tesselation. The asymptotic size complexity of the PSLG remains the same by this \lq convexification\rq \ operation. 

The \emph{Inverse Voronoi Problem} (IVP) consists of deciding whether $G$ coincides with the Voronoi diagram of some set $S$ of points in the plane, and if so, finding $S$. This problem was first studied by Ash and Bolker~\cite{ash-bolker}. Subsequently, Aurenhammer presented a more efficient algorithm~\cite{aurenhammer}, which in turn was improved by Hartvigsen, with the aid of linear programming ~\cite{hartvig92}, and later by Schoenberg, Ferguson and Li~\cite{schoen03}. Yeganova also used linear programming to determine the location of $S$~\cite{yega01,yeganova-thesis}. 

\begin{figure}[htbp]
\begin{center}
	 	\includegraphics[width=0.7\textwidth]{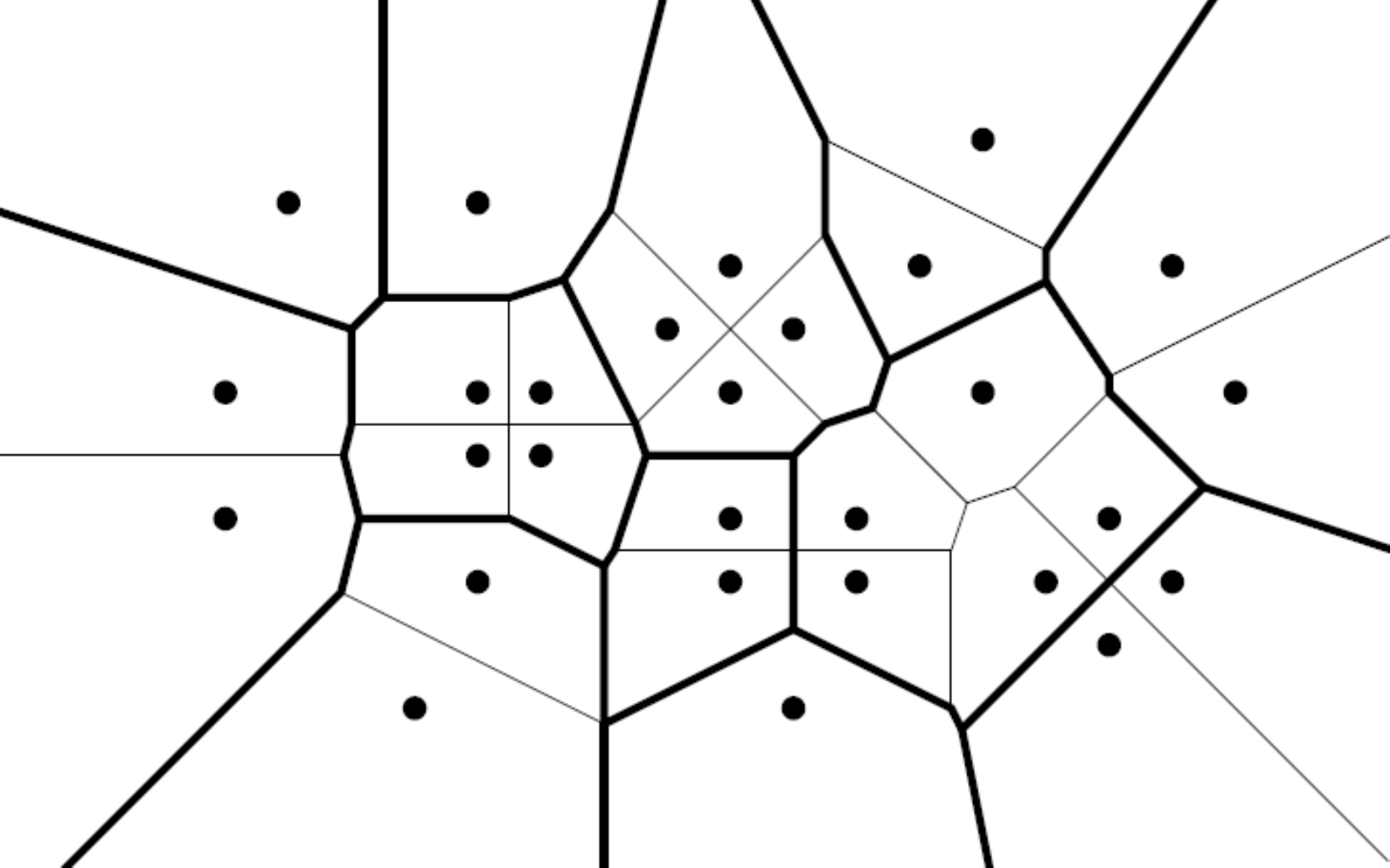}
	 \caption{GIVP: Thick edges represent the original input tesselation}
\end{center}	
	 \label{givp1}
\end{figure}

In the IVP, the set $S$ is limited to have one point per cell; a generalized version of this problem (GIVP) allows more than one point per cell. In this case, new vertices and edges may be added to $G$, but the original ones must be kept, as shown in Figure \ref{givp1}. With this relaxation the set $S$ always exists, hence we are interested in minimizing its size.

The GIVP in $\R^2$ was indirectly mentioned in \cite{yega01,yeganova-thesis}, in the context of set separation. It was formally stated and discussed in the III Cuban Workshop on Algorithms and Data Structures, held in Havana in 2003, where an algorithm for solving the problem in $\R^2$ was sketched by the current authors. However, the manuscript remained dormant for several years, and the algorithm was only published in Spanish in 2007 \cite{Trin07}. Recently, the problem was revisited in \cite{Baner12}, where another algorithm for the GIVP in $\R^2$ is given, and the special case of a rectangular tesselation is discussed in greater detail. The authors of \cite{Baner12} were unaware of \cite{Trin07}, however the two algorithms turn out to have certain common aspects. 

This paper is an expanded and updated English version of \cite{Trin07}. It contains a description and analysis of the aforementioned algorithm for solving the GIVP in $\R^2$. This is followed by the description of an implementation of the algorithm, which was used to make a first (if only preliminary) experimental study of the algorithm's performance. Our algorithm generates $\bigoh(E)$ sites in the worst case, where $E$ is the number of edges of $G$ (provided that the smallest angle of $G$ is constant). This bound is asymptotically optimal for tesselations with such angular constraints.

In comparison, the analysis given for the algorithm in~\cite{Baner12} states that  $\bigoh(V^3)$ sites are generated, 
where $V$ is the size of a refinement of $G$ such that all faces are triangles with acute angles. Given an arbitrary PSLG, there does not appear to be any known polynomial upper bound on the size of its associated acute triangulation. Even though it seems to us that the analysis in~\cite{Baner12} should have given a tighter upper bound in terms of $V$, even a linear bound would not make much of a difference, given that $V$ can be very large compared to the size of $G$. The analysis in~\cite{Baner12} is purely theoretical, so it would be interesting to perform an experimental study to shed some light on the algorithm's performance in practice.

This paper is organized as follows: In Section \ref{aljuarizmi} we describe the algorithm and discuss its correctness and performance. In Section \ref{implement} we derive some variants of the general strategy, and deal with several implementation issues of each variant. Section \ref{exper} is devoted to an experimental analysis of the algorithm's performance. Finally, in Section \ref{open} we summarize our results and discuss some open problems arising as a result of our work.

\section{The Algorithm}
\label{aljuarizmi}

First we establish some notation and definitions. In that respect we have followed some standard texts, such as \cite{deBerg}. 

Let $p$ and $q$ be points of the plane; as customary, $\overline{pq}$ is the segment that joins $p$ and $q$, and $\vert \overline{pq} \vert$ denotes its length. $B_{pq}$ denotes the bisector of $p$ and $q$, and $H_{pq}$ is the half-plane determined by $B_{pq}$, containing $p$. For a set $S$ of points in the plane, Vor($S$) denotes the \emph{Voronoi diagram} generated by $S$. The points in $S$ are called \emph{Voronoi sites} or \emph{generators}. 

If $p \in S$, $V(p)$ denotes the cell of Vor($S$) corresponding to the site $p$. For any point $q$, $C_S(q)$ is the largest empty circle centered at $q$, with respect to $S$  (the subscript $S$ can be dropped if it is clear from the context). Two points $p, q \in S$ are said to be \emph{(strong) neighbors} (with respect to $S$) if their cells share an edge in Vor($S$); in this case $E_{pq}$ denotes that edge. 

We will make frequent use of the following basic property of Voronoi diagrams:

\begin{lemma}[\cite{deBerg}, Thm. 7.4, p. 150]
\label{when-edge}
The bisector $B_{pq}$ defines an edge of the Voronoi diagram if, and only if, there exists a point $x$ on $B_{pq}$ such that $C_S(x)$ contains both $p$ and $q$ on its boundary, but no other site. The (open) edge in question consists of all points $x$ with that property. 
\end{lemma}

The technique used by the algorithm is to place pairs of points (\emph{sentinels}) along each edge $e$ of the PSLG (each pair is placed so that it is bisected by $e$) in order to \lq guard\rq \ or \lq protect\rq \ $e$. The number of sentinels required to protect $e$ depends on its length and the relative positions of its neighboring edges. Each pair of sentinels meant to guard $e$ is placed on the boundary of some circle, whose center lies on $e$. Furthermore this circle will not touch any other edge.  
The only exception is when the circle is centered on an endpoint of $e$, in which case it is allowed to touch all other edges sharing that endpoint.   More formally, we have the following.

\begin{definition}
Let $G$ be a PSLG, and let $e$ be an edge of $G$. Let $S$ be a set of points, and $p, q \in S$. The pair of points $p, q$ is said to be a \textbf{pair of sentinels} of $e$ if they are strong neighbors with respect to $S$, and $E_{pq}$ is a subsegment of $e$. In this case, $e$ (or more precisely, the segment $E_{pq}$) is said to be \textbf{guarded} by $p$ and $q$. 
\end{definition}

The algorithm works in two stages: First, for each vertex $v$ of $G$ we draw a circle centered on $v$. This is our set of \emph{initial circles} (this is described in more detail below). Then we proceed to cover each edge $e$ of $G$ by non-overlapping \emph{inner circles}, whose centers lie on $e$, and which do not intersect any other edges of $G$.

\begin{figure}[h!]
\begin{center}
	 	\includegraphics[width=0.6\textwidth]{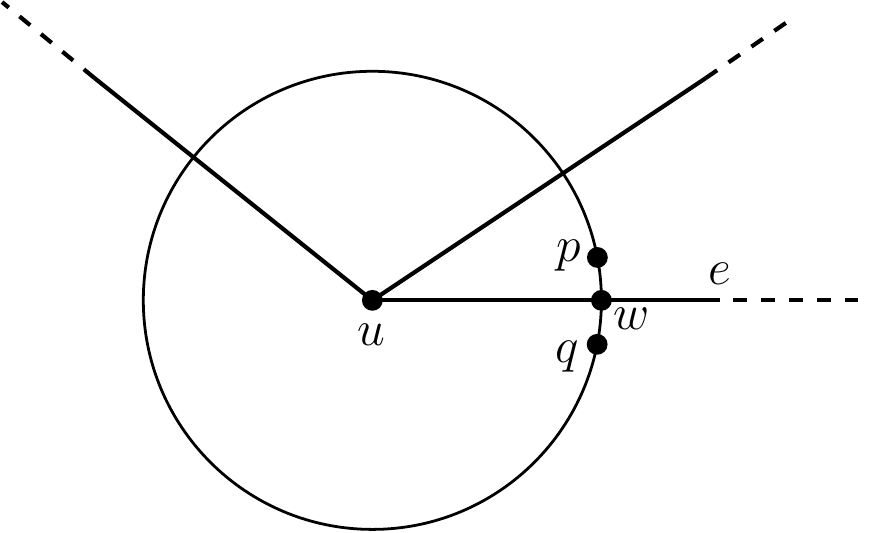}
	 \caption{Initial circle $\xi_G(u)$ for vertex $u$ and sentinels of $e$}
	 \label{circle1}
\end{center}	
\end{figure}

Let $u$ be a given vertex of $G$, and let $\lambda$ be the length of the shortest edge of $G$ incident to $u$. We denote as $\xi_G(u)$ the initial circle centered at $u$, which will be taken as the largest circle with radius $\rho_0 \leq \lambda / 2$ that does not intersect any edge of $G$, except those that are incident to $u$. Once we have drawn $\xi_G(u)$, for each edge $e$ incident to $u$ we can choose a pair of sentinels $p, q$, placed on $\xi_G(u)$, one on each side of $e$, at a suitably small distance $\epsilon$ from $e$, as in Figure \ref{circle1}.   Later in this section we discuss how to choose $\epsilon$ appropriately.

Let $w$ be the point of intersection between $\xi_G(u)$ and $e$; now $p$ and $q$ guard the segment $\overline{uw}$ of $e$, which means that $\overline{uw}$ will appear in the Voronoi diagram that will be constructed, provided that we do not include any new points inside $\xi_G(u)$ (see Lemma \ref{when-edge}). 

\begin{figure}[htbp]
\begin{center}
	 	\includegraphics[width=0.9\textwidth]{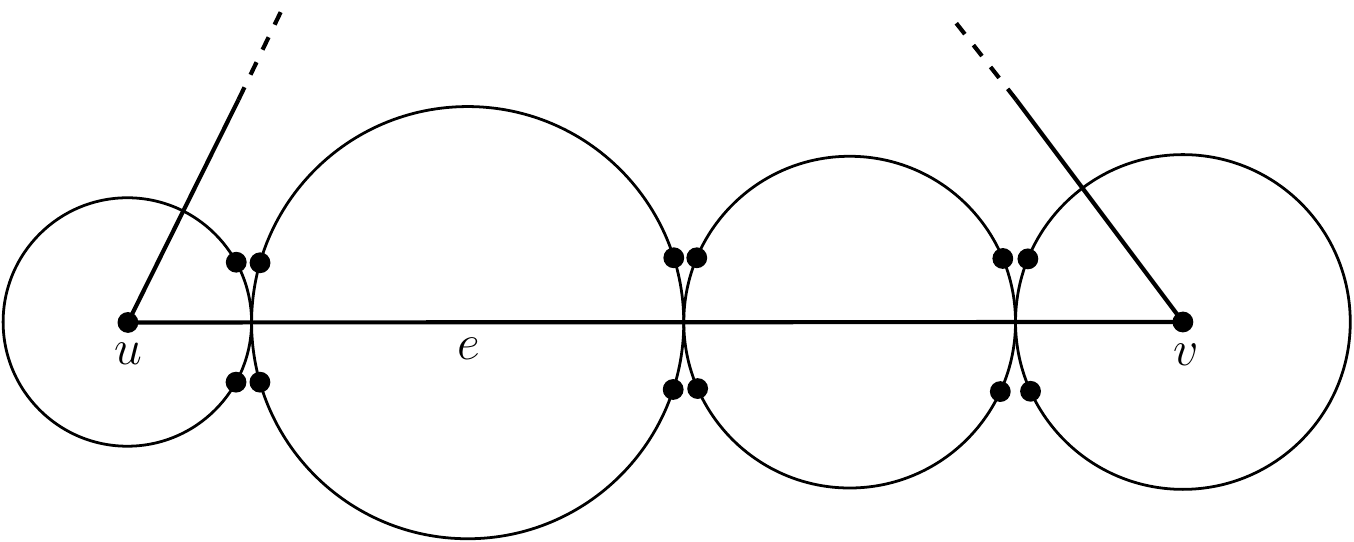}
	 \caption{Edge covered by circles}
	 \label{circle2}
\end{center}	
\end{figure}

Let $e = \overline{uv}$ be an edge of $G$, and $w_1, w_2$ the intersection points of $\xi(u)$ and $\xi(v)$ with $e$, respectively.\footnote{For convenience, we have dropped the subscript $G$.} The segments $\overline{u w_1}$ and $\overline{u w_2}$ are now guarded, whereas the (possibly empty) segment $\overline{w_1 w_2}$ still remains unguarded. In order to guard $\overline{w_1 w_2}$ it suffices to cover that segment with circles centered on it, not intersecting with any edge other than $e$, and not including any sentinel belonging to another circle. Then we can choose pairs of sentinels on each covering circle, each sentinel being at distance $\epsilon$ from $e$, as shown in Figure \ref{circle2}. 

\noindent As a consequence of Lemma \ref{when-edge}, $e$ will be guarded in all its length, provided that no new point is later included inside one of the circles centered on $e$.  To ensure this, we will not allow an inner circle of $e$ to get closer than $\epsilon$ to another edge $f$, because then a sentinel of $f$ might fall inside the circle. With this precaution, the sentinels guarding $e$ will not interfere with other edges, since they will not be included in any circle belonging to another edge. 

In summary, an outline of the algorithm is:

\begin{enumerate}
\item For each vertex $u \in G$, draw initial circle $\xi_G(u)$ centered on $u$. 
\item Choose a suitable value of $\epsilon$.
\item For each vertex $u$ and for each edge $e$ incident to $u$, place a pair of sentinels on $\xi_G(u)$, symmetric to one another with respect to $e$, at distance $\epsilon$ from $e$.
\item For each edge $e \in G$, cover the unguarded segment of $e$ with inner circles centered on $e$, and then place pairs of sentinels on each circle. 
\end{enumerate}

\noindent This algorithm is a general strategy that leads to several variants when Step 4 is specified in more detail, as will be seen in Section \ref{implement}. In order to prove that the algorithm works it suffices to show that: 

\begin{enumerate}
\item The algorithm terminates after constructing a finite number of circles (and sentinels).
\item After termination, every edge of $G$ is guarded (see the discussion above). 
\end{enumerate}

\noindent In order to show that the algorithm terminates we will establish some facts. Let $\rho_0>0$ be the radius of the smallest initial circle.
Now let $\alpha$ be the smallest angle formed by any two incident edges of $G$, say $e$ and $f$. 
By taking $\epsilon \leq \rho_0 \sin \frac{\alpha}{2}$ we make sure that
any sentinel will be closer to the edge that it is meant to guard than to any other edge. This is valid for all initial circles. 

\begin{figure}[h!]
\begin{center}
	 	\includegraphics[width=0.8\textwidth]{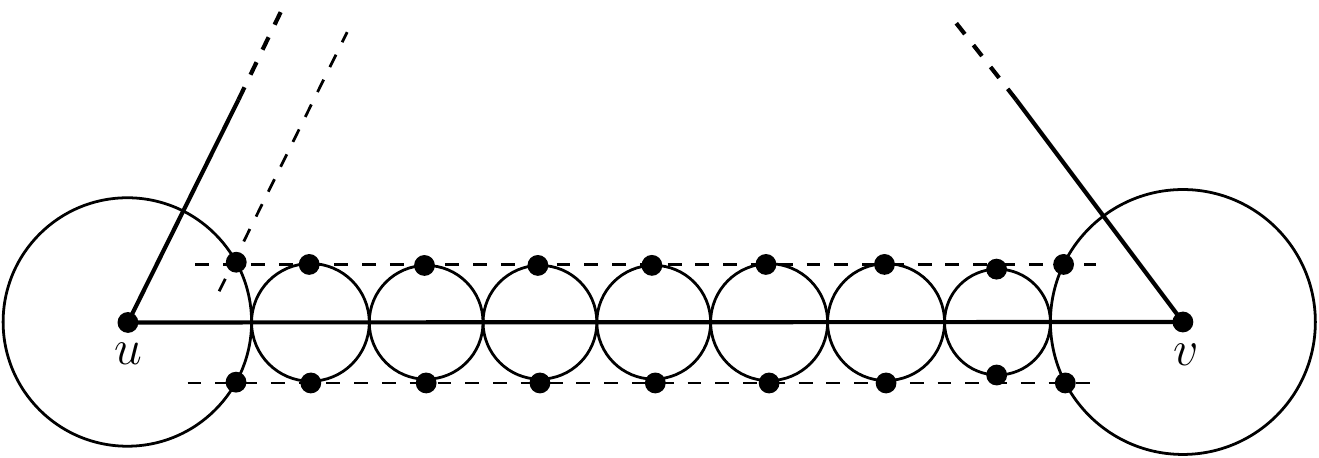}
	 \caption{Covering the middle segment of edge $\overline{uv}$ by inner circles of radius $\epsilon$}
	 \label{circle3}
\end{center}	
\end{figure}

After all initial circles have been constructed, together with their corresponding sets of sentinels, for every edge $e$ there may be a {\em middle segment} that remains unguarded. This segment must be covered by a finite number of inner circles. Take one edge, say $e$, with middle unguarded segment of length $\delta$. If we use circles of radius $\epsilon$ to cover the unguarded segment, then we can be sure that these circles will not intersect any circle belonging to another edge. Exactly $\lfloor \delta / 2\epsilon \rfloor +1$ such circles will suffice to cover the middle segment, where the last one may have a radius $\epsilon'$ smaller than $\epsilon$. For this last circle, the sentinels could be placed at distance $\epsilon' < \epsilon$ from $e$ (c.f. Figure \ref{circle3}). 

Using circles of radius $\epsilon$ is, among all the possible variants mentioned here, the one that yields the largest number of circles, and hence the largest number of sentinels (generators of the Voronoi diagram). Now let $e$ be the longest edge of $G$, with length $\Delta$. In the worst case, the number of inner circles that cover $e$ will be $\lfloor (\Delta-2\rho_0) / 2\epsilon \rfloor +1$, and the number of sentinels will be twice that number plus four (corresponding to the sentinels of both initial circles). Therefore, the algorithm generates a number of points that is linear in $E$, the number of edges, which is asymptotically optimal, since a lower bound for the number of points is the number of faces in $G$. 

Note that by letting $G$ become part of the problem instance, the number of generators becomes a function of $\alpha$, and it is no longer linear in $E$. In practice, however, screen resolution and computer arithmetic impose lower bounds on $\alpha$. Under such constraints, the above analysis remains valid. This leads to our main result: 

\begin{theorem}
Let $G$ be a planar straight-line graph, whose smallest angle $\alpha$ is larger than a fixed constant. Then, the corresponding Generalized Inverse Voronoi Problem can be solved with $\bigoh(E)$ generators, where $E$ is the number of edges of $G$.  
\end{theorem}

\section{Implementation}
\label{implement}

In step 4 of the algorithm given in the previous section, the method to construct the inner circles was left unspecified. Taking the circles with radius $\epsilon$, as suggested in the preceding analysis, is essentially a brute-force approach, and may easily result in too many sentinels being used. In this section we discuss two different methods for constructing the inner circles. 

\begin{figure}[h!]
\begin{center}
	 	\includegraphics[width=0.8\textwidth]{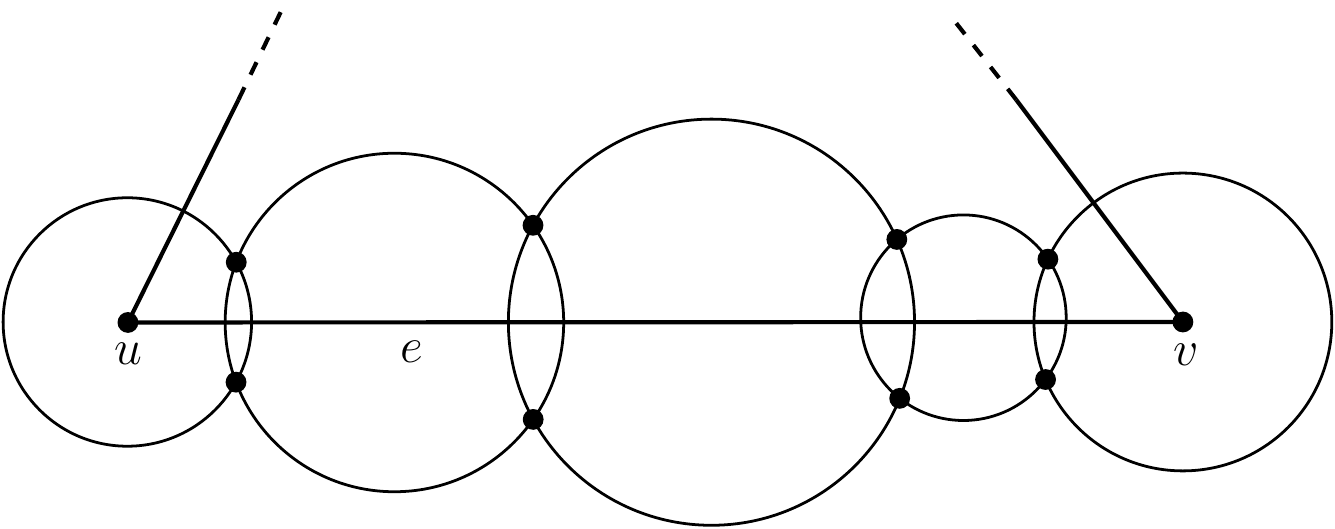}
	 \caption{Adjacent circles share a pair of sentinels}
	 \label{circle2B}
\end{center}	
\end{figure}

First let us note that in order to reduce the number of sentinels in our construction we may allow two adjacent circles on the same edge to overlap a little, so that they can share a pair of sentinels (see Figure \ref{circle2B}). This observation is valid for all variants of the algorithm. 

The first variant for the construction of the inner circles along an edge is to place them sequentially (iteratively), letting them grow as much as possible, provided that they do not enter the $\epsilon$-wide \lq security area\rq \ of another edge. Obviously, this greedy heuristic must yield a smaller number of Voronoi generators than the naive approach of taking all circles with radius $\epsilon$. 

Suppose we want to construct an inner circle $\chi$ for  edge $e$, adjacent to another circle on $e$ that has already been fixed and on which we have already placed two sentinels: $a=(x_a, y_a)$, and $b=(x_b, y_b)$. 
Let $f$ be the first edge that will be touched by $\chi$ as it grows, while constrained to have its center on $e$ and  $a,b$ on its boundary. 
Let $f'$ be a straight line parallel to $f$, at distance $\epsilon$ from $f$, and closer to $\chi$ than $f$. Let $e$ be defined by the equation $y=mx+n$, and $f'$ by the equation $Ax+By+C = 0$.\footnote{The equation of $f'$ can be obtained easily after the initial circles have been constructed and their sentinels placed.} The distance of any point $(x, y)$ to $f'$ is given by $\frac{ \vert Ax+By+C \vert }{\sqrt{A^2+B^2}}$. The radius of $\chi$ must be equal to this distance.
Hence the $x$-coordinate of the center satisfies the following quadratic equation: 
\[
(A^2+B^2)((x_a - x)^2+(y_a -(m x + n))^2)=(A x + B(m x + n) + C)^2
\]

or 

\[ -(A^2 + 2ABm + B^2m^2 - D(m^2+1))x^2 \]
\[ -2(A(Bn+C) + B^2mn + BCm + D(x_a - m(n-y_a )))x \]
\[ -B^2n^2 - 2BCn - C^2 + D(n^2-2ny_a +x_a ^2+y_a ^2)=0 \]

where $D=A^2+B^2$.
\\\\
Our second variant for constructing  inner circles is also based on the  principle of letting them grow until they come within distance $\epsilon$ of some edge. Yet, instead of growing the circles sequentially along the edge that is to be covered, we center the first inner circle on the midpoint of the unguarded middle segment.  This will yield at most two smaller disjoint unguarded segments, on which we recurse. In the worst case, a branch of the recursion will end when an unguarded segment can be covered by a circle of radius $\epsilon$.
The advantage of this approach is that the coordinates of the center can be determined with much less computation, thus avoiding potential roundoff errors. Additionally, this variant is more suitable for parallel implementation than the previous one. On the other hand, we need an extra data structure to handle the unguarded segments.   

We end this discussion with a word about the choice of $\epsilon$. On one hand, $\epsilon$ must be sufficiently small for the construction to be carried out. On the other hand, for the sake of robustness to numerical errors, it is convenient to take $\epsilon$ as large as possible. That is why we defer the actual choice of $\epsilon$ until the initial circles have been drawn. A different approach might be to use a variable-sized $\epsilon$, which would lead to a more complicated, yet (hopefully) more robust algorithm. 

A final remark: For the sake of simplicity we have assumed throughout the whole discussion that the cells of the input tesselation are convex, but our algorithm could be easily generalized to accept tesselations with non-convex cells. 

\section{Experimental Analysis}
\label{exper}

From the analysis in Section \ref{aljuarizmi} we know that the number of sites generated by our algorithm is linear in the size of the input, provided that the smallest angle $\alpha$ is constant. However, we would like to get a more precise idea about the algorithm's performance, and the difference between the two strategies we have suggested for Step 4. For that purpose, we have implemented the algorithm and carried out a set of experiments. 

Our experimental workbench consists of a Graphical User Interface, which can generate a tesselation on a random point set, store it in a DCEL data structure, and then apply one of the two variants of the algorithm for solving the GIVP, described in Section \ref{aljuarizmi}.

The GUI is described in more detail in \cite{Trin05,Trin07}, and a beta Windows version can be downloaded from  \url{https://www.researchgate.net/publication/239994361_Voronoi_data}. The file \lq Voronoi data.rar\rq \ contains the Windows executable and a few DCEL files, consisting of sample tesselations. The user can generate additional tesselations randomly, and apply either variant of the algorithm on them. 

The tesselations are generated as follows: First, the vertex set of $G$ is randomly generated from the uniform distribution in a rectangular region. Then, pairs of vertices are chosen randomly to create edges. If a new edge intersects existing edges, then the intersection points are added as new vertices, and the intersecting edges are decomposed into their non-intersecting segments. Finally, some edges are added to connect disjoint connected components and dangling vertices, so as to make the PSLG biconnected. 

Table \ref{tab:results} displays some statistics about 40 such randomly generated tesselations: Number of vertices, number of edges, number of regions, number of Voronoi sites with the recursive version of Step 4, number of Voronoi sites with the sequential version of Step 4, the smallest angle $\alpha$, and the width $\epsilon$ of the security area. The tesselations have been listed in increasing order of the number of edges. For each parameter, the table also provides the median (MED), the mean value (AVG), and the standard deviation (STD). 

\begin{table}[htp]
\begin{tabular}{|cc|*{3}{c}|*{2}{c}|c|c|} \hline
\multicolumn{2}{|c|}{} & \multicolumn{3}{c|}{} & \multicolumn{2}{c|}{\textbf{Num. of sites generated}} &  &  \\
\multicolumn{2}{|c|}{\textbf{Exp.}} &  \multicolumn{3}{c|}{\textbf{Tesselation}}  &  &  & \textbf{Smallest} &  \\
\multicolumn{2}{|c|}{\textbf{num.}} &  &  &  & \textbf{Recursive} & \textbf{Sequential} & \textbf{angle} &  $\epsilon$\textbf{-neigh.} \\
\multicolumn{2}{|c|}{} & \textbf{Vertices} & \textbf{Edges} & \textbf{Regions} & \textbf{version} & \textbf{version} & \textbf{(degrees)} & \textbf{(pixels)} \\ 
\hline  
\hline 
\multicolumn{2}{|c|}{1} & 72 & 142 & 66 & 1 020 & 852 & 1.63 & 1.20 \\ 
\multicolumn{2}{|c|}{2} & 117 & 206 & 91 & 916 & 870 & 4.09 & 12.30 \\ 
\multicolumn{2}{|c|}{3} & 194 & 252 & 60 & 1 468 & 1 296 & 1.07 & 9.61 \\ 
\multicolumn{2}{|c|}{4} & 274 & 376 & 105 & 1 672 & 1 596 & 1.60 & 12.26 \\
\multicolumn{2}{|c|}{5} & 229 & 429 & 202 & 2 400 & 2 148 & 3.38 & 0.91 \\ 
\multicolumn{2}{|c|}{6} & 314 & 441 & 129 & 2 208 & 2 020 & 0.56 & 5.70 \\ 
\multicolumn{2}{|c|}{7} & 336 & 472 & 138 & 2 656 & 2 374 & 0.47 & 4.03 \\ 
\multicolumn{2}{|c|}{8} & 339 & 475 & 138 & 3 098 & 2 618 & 3.95 & 0.18 \\
\multicolumn{2}{|c|}{9} & 344 & 480 & 138 & 3 140 & 2 720 & 0.23 & 4.48 \\
\multicolumn{2}{|c|}{10} & 357 & 493 & 138 & 2 844 & 2 530 & 0.13 & 7.24 \\
\multicolumn{2}{|c|}{11} & 339 & 501 & 164 & 2 580 & 2 364 & 0.38 & 3.81 \\ 
\multicolumn{2}{|c|}{12} & 390 & 568 & 180 & 2 680 & 2 520 & 8.92 & 0.60 \\
\multicolumn{2}{|c|}{13} & 438 & 637 & 281 & 3 320 & 3 028 & 0.21 & 6.34 \\
\multicolumn{2}{|c|}{14} & 403 & 641 & 240 & 3 838 & 3 382 & 0.16 & 7.07 \\
\multicolumn{2}{|c|}{15} & 472 & 684 & 214 & 3 432 & 3 144 & 0.25 & 2.56 \\
\multicolumn{2}{|c|}{16} & 397 & 721 & 319 & 4 244 & 3 718 & 0.11 & 3.16 \\
\multicolumn{2}{|c|}{17} & 421 & 784 & 365 & 5 112 & 4 406 & 1.30 & 0.12 \\
\multicolumn{2}{|c|}{18} & 564 & 826 & 264 & 4 092 & 3 790 & 0.70 & 0.11 \\
\multicolumn{2}{|c|}{19} & 504 & 986 & 463 & 4 148 & 4 020 & 2.37 & 14.16 \\
\multicolumn{2}{|c|}{20} & 512 & 999 & 472 & 4 276 & 4 134 & 1.52 & 3.68 \\
\multicolumn{2}{|c|}{21} & 552 & 1 056 & 506 & 4 048 & 4 796 & 0.25 & 4.40 \\
\multicolumn{2}{|c|}{22} & 574 & 1 107 & 535 & 4 856 & 4 689 & 3.68 & 0.75 \\
\multicolumn{2}{|c|}{23} & 601 & 1 166 & 567 & 5 240 & 5 009 & 0.80 & 3.43 \\
\multicolumn{2}{|c|}{24} & 645 & 1 256 & 613 & 5 852 & 5 521 & 0.77 & 3.62 \\
\multicolumn{2}{|c|}{25} & 672 & 1 292 & 622 & 5 720 & 5 992 & 0.25 & 3.44 \\
\multicolumn{2}{|c|}{26} & 738 & 1 311 & 575 & 6 124 & 5 832 & 0.34 & 1.84 \\
\multicolumn{2}{|c|}{27} & 724 & 1 399 & 677 & 6 440 & 6 194 & 0.23 & 3.23 \\
\multicolumn{2}{|c|}{28} & 815 & 1 441 & 628 & 6 832 & 6 478 & 1.20 & 0.30 \\
\multicolumn{2}{|c|}{29} & 763 & 1 479 & 718 & 6 960 & 6 599 & 2.14 & 0.19 \\
\multicolumn{2}{|c|}{30} & 772 & 1 495 & 725 & 6 900 & 6 610 & 2.54 & 0.29 \\
\multicolumn{2}{|c|}{31} & 855 & 1 522 & 669 & 7 684 & 7 158 & 1.43 & 0.31 \\ 
\multicolumn{2}{|c|}{32} & 894 & 1 607 & 712 & 9 062 & 8 685 & 0.36 & 0.23 \\ 
\multicolumn{2}{|c|}{33} & 898 & 1 615 & 716 & 8 152 & 7 580 & 1.19 & 0.33 \\ 
\multicolumn{2}{|c|}{34} & 963 & 1 750 & 789 & 9 637 & 9 045 & 0.88 & 0.29 \\ 
\multicolumn{2}{|c|}{35} & 1 006 & 1 842 & 838 & 9 236 & 8 582 & 1.85 & 0.34 \\
\multicolumn{2}{|c|}{36} & 1 018 & 1 874 & 858 & 10 144 & 9 228 & 1.09 & 0.27 \\
\multicolumn{2}{|c|}{37} & 984 & 1 902 & 920 & 7 924 & 7 792 & 4.40 & 0.25 \\
\multicolumn{2}{|c|}{38} & 1 015 & 1 962 & 949 & 8 396 & 8 198 & 3.34 & 0.31 \\
\multicolumn{2}{|c|}{39} & 1 066 & 1 973 & 909 & 10 392 & 9 492 & 0.63 & 0.30 \\
\multicolumn{2}{|c|}{40} & 1 019 & 1 999 & 982 & 8 952 & 8 616 & 1.49 & 0.45 \\
\hline
\multicolumn{2}{|c|}{\textbf{MED}} & \textbf{558} & \textbf{1 027.5} & \textbf{489} & \textbf{4 566} & \textbf{4 547.5} & \textbf{3.4} & \textbf{0.3} \\
\multicolumn{2}{|c|}{\textbf{AVG}} & \textbf{590} & \textbf{1 054} & \textbf{467} & \textbf{5 192} & \textbf{4 891} & \textbf{0.59} & \textbf{4.06} \\
\multicolumn{2}{|c|}{\textbf{STD}} & \textbf{282.7} & \textbf{571.24} & \textbf{292.73} & \textbf{2 715.5} & \textbf{2 600} & \textbf{3.36} & \textbf{0.74} \\
\hline
\end{tabular}
\vspace{5mm}
\caption{Experimental results}
\label{tab:results}
\end{table}

From the tabulated data we can also get empirical estimates about the correlation among different parameters, especially $\alpha$ and $\epsilon$, and about the distribution of their values. The parameters $\alpha$ and $\epsilon$ show a weak negative correlation with the number of edges, of $-0.548$ and $-0.358$ respectively. In turn they are positively correlated with one another, with a correlation of $0.67$. These empirical findings agree with intuition. 

\begin{figure}[htbp]
\begin{minipage}{0.5\linewidth}
\centering
\includegraphics[width=1.0\textwidth]{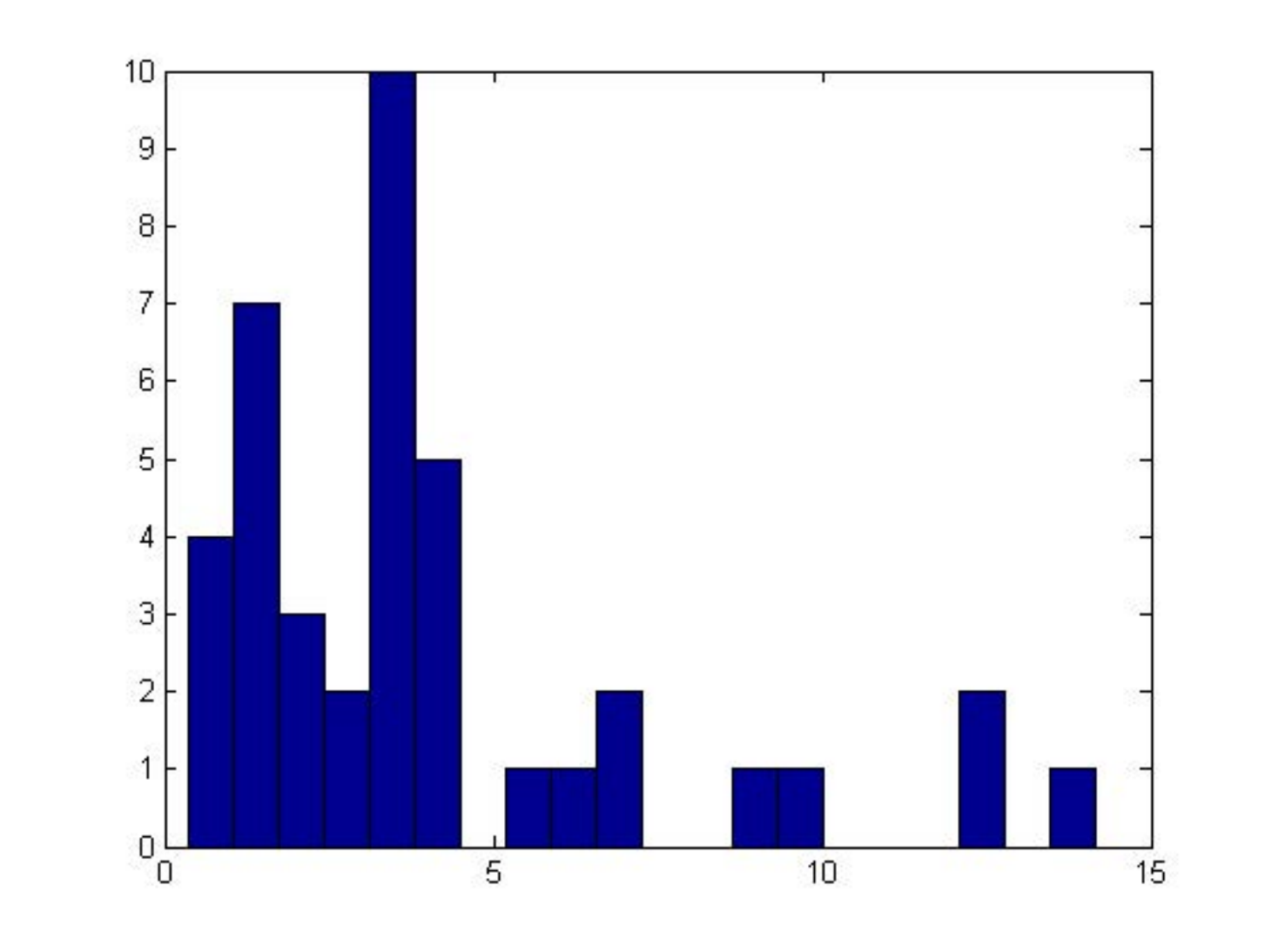}
\caption{Histogram of $\alpha$}
\label{histoalpha}
\end{minipage}
\begin{minipage}{0.5\linewidth}
\centering
\includegraphics[width=1.0\textwidth]{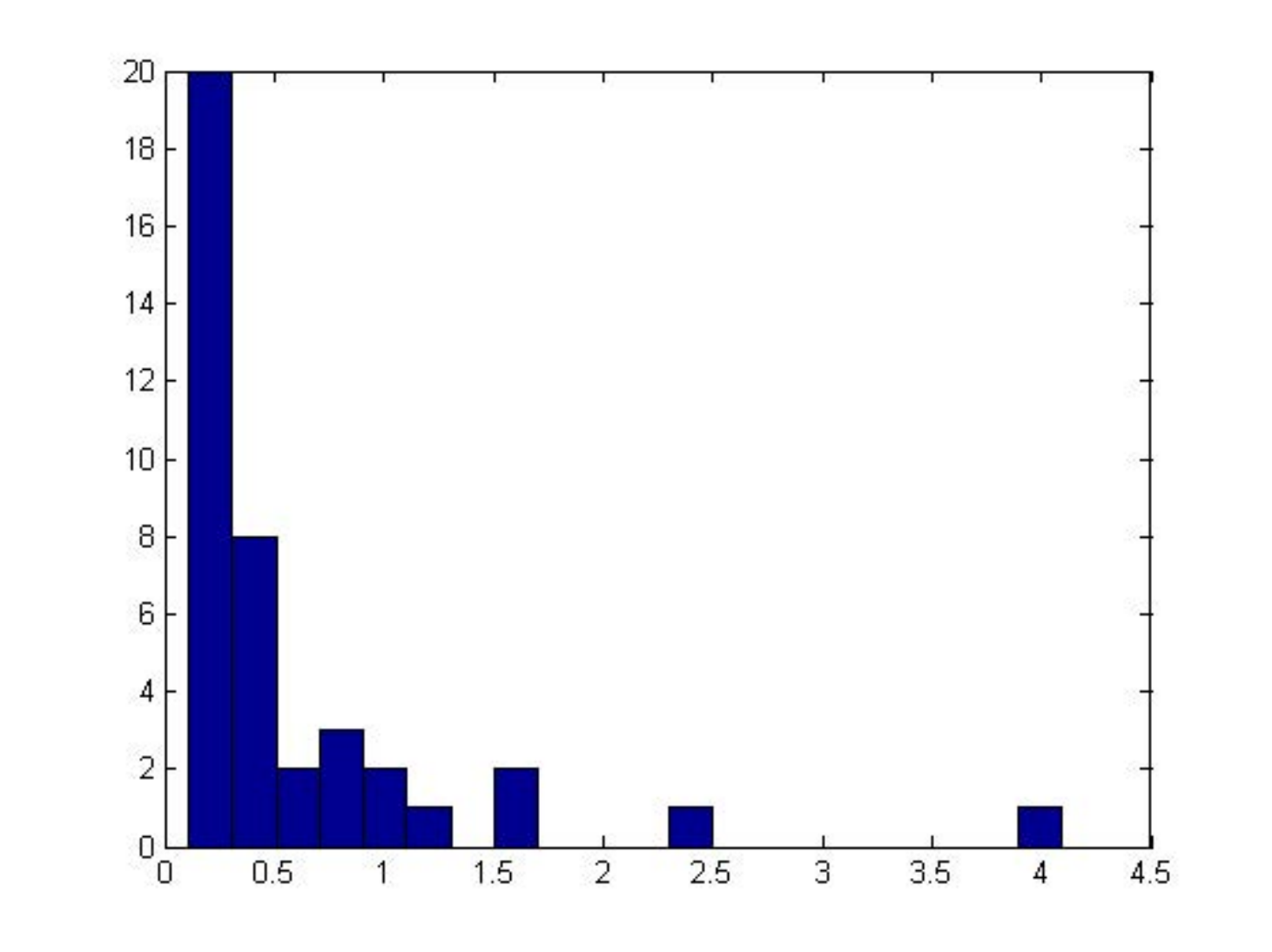}
\caption{Histogram of $\epsilon$}
\label{histoepsilon}
\end{minipage}
\end{figure}

Figures \ref{histoalpha} and \ref{histoepsilon} display the histograms of $\alpha$ and $\epsilon$ with 20 bins. They can be well approximated by Poisson distributions, with $\lambda = 4.06$ and $\lambda = 0.59$, respectively, and with 95\% confidence intervals $[3.437; 4.686]$ and $[0.375; 0.8784]$. 

\begin{figure}[htbp]
	 	\includegraphics[width=1.8\textwidth, angle=90]{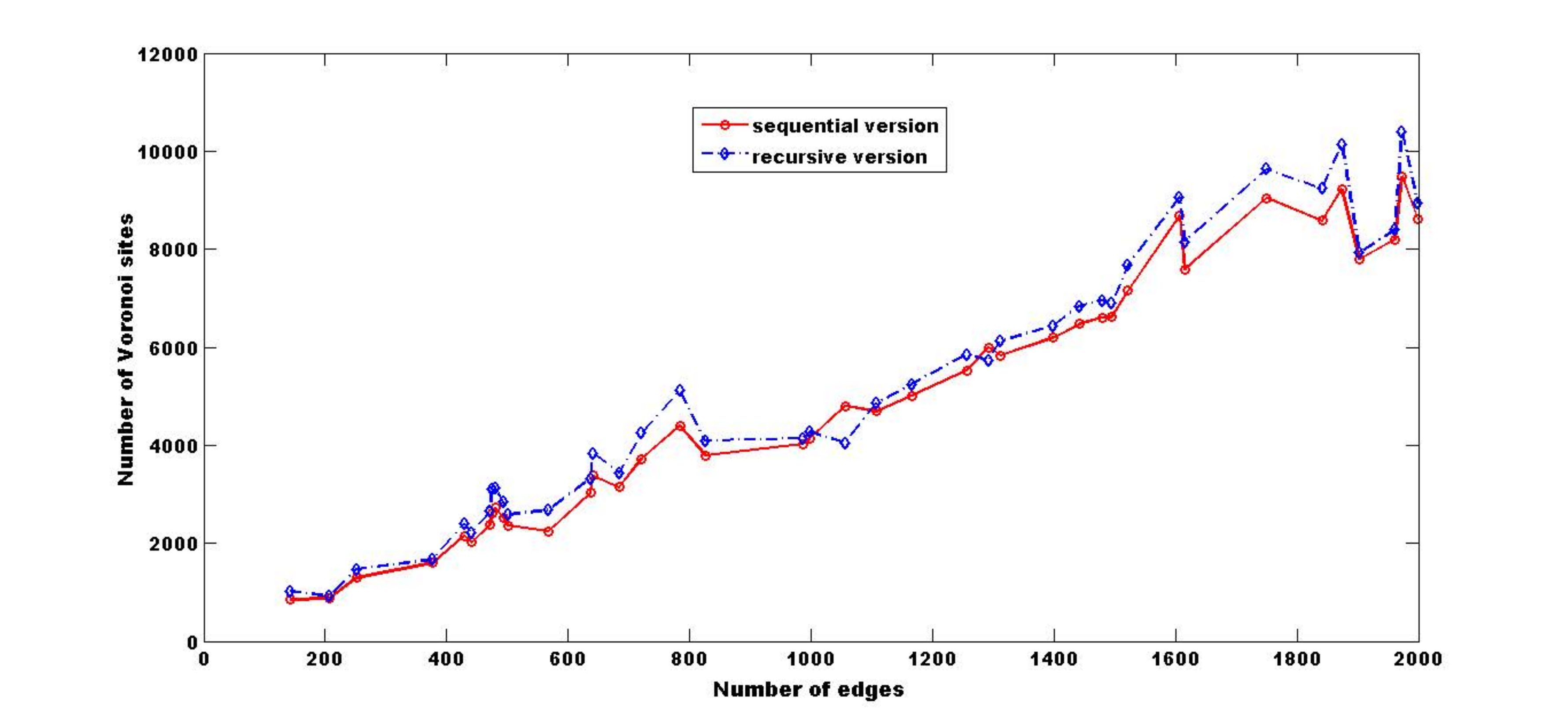} 
	 \caption{Plot of the results in Table \ref{tab:results}}
	 \label{plot1B}
\end{figure}

The comparison between the two variants of the algorithm is shown in Figure \ref{plot1B}. We can see that the sequential variant is slightly better than the recursive variant, as it generates a smaller number of sites in most cases. However, the difference between both variants is not significant. Indeed, the linear regression fits have very similar slopes: The linear fit for the sequential variant is $y = 4.4831 x + 158.59$, whereas the linear fit in the recursive case is $y = 4.6241 x + 318.51$. 

\section{Conclusions and Open Problems}
\label{open}

Our results show that the Generalized Inverse Voronoi Problem can be solved with a number of generators that is linear in the size of the input tesselation, provided that we enforce a lower bound on the size of the smallest angle. On the other hand, the algorithm described in \cite{Baner12} produces $\bigoh(V^3)$ generators, where $V$ is the number of vertices of an acute triangulation of $G$. As the performance of the two algorithms is given as a function of different parameters, a theoretical comparison between them is not straightforward.  
%
%
An experimental study could be helpful, but that would require an implementation of the algorithm in \cite{Baner12}. In practice, our algorithm generates approximately $4.48E + 159$ Voronoi sites, where $E$ is the number of edges of the input tesselation. 

In any case, the number of generators produced by both algorithms may still be too large, and it may be possible to reduce it to a number closer to $F$, the number of faces of the tesselation, which is the trivial lower bound. This lower bound can only be achieved if the tesselation is a Voronoi tesselation. In the more general case, how close to $F$ can we get? 

In particular, our algorithm still has plenty of room for improvement. In Section \ref{implement} we have already mentioned several strategies that can decrease the number of Voronoi sites produced. The design of a parallel version, and a version that is robust against degenerate cases and numerical roundoff errors, are other issues to consider. Roundoff errors  have long been an important concern in Computational Geometry in general, and in Voronoi diagram computation, in particular (see e.g. \cite{sugi92}). 

Other practical questions have to do with the experimental analysis of our algorithms. We have devised a method to generate a PSLG on a random point set, but we have not analyzed how this compares to generating such graphs uniformly from the set of all PSLGs that can be defined on a given point set. Regarding certain properties of our generated graphs (expected number of vertices, edges, and faces, expected area of the faces, distribution of the smallest angle, etc.), we have not attempted a theoretical analysis, but we have estimated some of these parameters empirically. A more comprehensive set of experiments will reveal how these tesselations compare with those generated by other methods. 

As a final remark, we point out that our algorithm could also be generalized to other metrics, continuous or discrete, including graph metrics. Potential applications include image representation and compression, as described in \cite{Mar07}, and pattern recognition (e.g. given a partition of some sample space, we could select a set of \lq representatives\rq \ for each class). In the case of graphs, Voronoi partitions can be used to find approximate shortest paths (see \cite{Som10,Ra12}, for instance). In social networks, node clustering around a set of \lq representative\rq \ nodes, or \lq super-vertices\rq, is a popular technique for network visualization and$/$or anonymization \cite{Zhou08}. 

\section*{Acknowledgements}

Janos Pach contributed some key ideas for the algorithm, at the early stages of this work. Hebert P\'erez-Ros\'es was partially supported by the Spanish Ministry of Economy and Competitiveness, under project TIN2010-18978. Guillermo Pineda-Villavicencio was supported by a postdoctoral fellowship funded by the Skirball Foundation, via the Center for Advanced Studies in Mathematics at the Ben-Gurion University of the Negev, Israel, and by an ISF grant.


\end{document}